\documentclass[a4paper]{article}
\usepackage{graphics,bm,amsmath}

\newcommand{\ket}[2]{|#1\rangle _{#2}}
\newcommand{\bra}[2]{\langle _{#1}#2|}
\newcommand{\up}{\!\uparrow}
\newcommand{\down}{\!\downarrow}

\title{Proper and Improper Separability}
\author{Christopher G. Timpson\thanks{\texttt{c.g.timpson@leeds.ac.uk}}\\ \textit{{\small Division of History and Philosophy of Science,}}\\ \textit{ {\small School of Philosophy, University of Leeds, LS2 9JT, UK.}} \and Harvey R. Brown\thanks{\texttt{harvey.brown@philosophy.ox.ac.uk}}\\ \textit{\small Faculty of Philosophy,}\\ \textit{\small University of Oxford, 10 Merton St, Oxford, OX1 4JJ, UK.}}
\date{23 May 2005}

\begin{document}
\maketitle

\begin{abstract}
\noindent The distinction between proper and improper mixtures is a staple of the discussion of foundational questions in quantum mechanics. Here we note an analogous distinction in the context of the theory of entanglement. The terminology of `proper' versus `improper' separability is proposed to mark the distinction.
\end{abstract}

\section{Proper and Improper mixtures}
In many discussions of the measurement problem in quantum mechanics, it has proved essential to distinguish between density operators which can be given an `ignorance' interpretation; and those which cannot. 
In the former case, the system whose state the density operator represents is in some definite quantum state from a specified set --- say a pure state $\ket{\psi_{i}}{}$ --- but we don't know which. Our ignorance of the actual state can be represented by a probability distribution $\{p_{j}\}$ over the different (not necessarily orthogonal) possibilities $\{\ket{\psi_{j}}{}\}$, and the density operator may be written as
\begin{equation} 
\rho=\sum_{j}p_{j}\ket{\psi_{j}}{}\bra{}{\psi_{j}}. \label{proper}
\end{equation}
An ensemble of such systems will also be described by the density operator (\ref{proper}); in this case each of the elements of the ensemble will be in a definite state $\ket{\psi_{i}}{}$, and the relative frequency with which elements of the ensemble are in the $j$-th state will be given by the probability $p_{j}$. One usually thinks of such an ensemble as arising from a preparation procedure which, on any given run, produces the state $\ket{\psi_{j}}{}$ with probability $p_{j}$. Ensembles which can be given an ignorance interpetation are called \textit{proper} mixtures; the terminology is due to d'Espagnat \cite{d'Espagnat}.
While the statistics that one will obtain following measurements on an ensemble of systems in a proper mixture will simply be governed by the density operator $\rho$, which, as a mathematical object, may be expressed in various forms \cite{schrodinger2,jaynes:1957ii,hjw}, it is nevertheless the case that there is only one privileged decomposition in the form (\ref{proper}) for a given proper mixture---that one expressing the actual probabilities and definite states according to which the systems have been mixed. It then follows, of course, that one will not be able to determine the actual composition of a proper mixture by measurements on the systems making up the ensemble alone, but will require some further information.  

By contrast, density operators which cannot be given an ignorance interpretation are said to represent \textit{improper} mixtures. Here the density operator arises from tracing out irrelevant, or unavailable, degrees of freedom. In this case, individual systems cannot be thought to be in some definite state of which we are ignorant; rather, the (reduced) density operator is the only description that they can have. The well-known ambiguity of the representation of a density operator now achieves free rein: as we have said, for a given operator $\rho$, there exist many decompositions of the form (\ref{proper}). For improper mixtures there is no longer any fact of the matter as to which decomposition is privileged or correct. Indeed the very notion of correctness has no application in this case.

The significance of the distinction between proper and improper mixtures for the measurement problem in orthodox quantum theory is, of course, just this: Proceeding within the orthodox theory, what one wants as the end result of a measurement procedure is to be left with a proper mixture, corresponding to definite measurement outcomes, whereas what one actually gets (in the absence of the problematic process of genuine collapse) is an improper mixture; even if it is one in which decoherence has ensured that off-diagonal elements of the density matrix in the position basis tend rapidly to zero \cite{decoherence}. And improper mixtures do not correspond to definite measurement outcomes. (That is, unless one abandons orthodoxy and relativizes the notion of definite outcome, leading to an Everett-type theory \cite{everett,simonrelativism}.)\footnote{For a recent exchange which betrays that there can still be some confusion over the notions of proper and improper mixtures, see \cite{Kirkpatrick} and d'Espagnat's reply \cite{d'Espagnatreply}.}       

At first sight, the distinction between proper and improper mixtures appears to suffer from a certain restriction in its range of application. According to interpretations in which quantum mechanics \textit{without} collapse is a complete theory, as in the Everett interpretation, a state being mixed is \textit{always} the result of tracing out unwanted degrees of freedom\footnote{The only exception would be if the universal state were itself mixed, rather than pure, as usually assumed.}. In this case, there would seem to be no scope left for the category of proper mixtures. However, even in no-collapse quantum mechanics, a useful distinction may be made between proper and improper mixtures, if this characterisation is recognised as relative to the experimental context.    

To see this, consider again the example of states $\ket{\psi_{j}}{}$ being produced with probabilities $p_{j}$.\footnote{The following sketch can be generalized to include cases in which mixtures of \textit{mixed} states are prepared.} In a no-collapse framework, we may model the preparation procedure, schematically, in the following way.

We begin with a preparation device, that is, some apparatus which, on any given run of the procedure, will prepare an object system in one of the states $\ket{\psi_{j}}{}$, depending upon the setting of some internal degree of freedom --- imagine a knob which may be turned to one of $j=1,\ldots,n$ different positions. Next, mixing probabilities are introduced. These probabilities arise because the setting of the knob for each run of the apparatus is controlled by some further degree of freedom, which we may think of as a quantum `die' that is thrown. This die has a set of orthogonal states $\{\ket{d_{j}}{}\}, j=1,\ldots,n$, which correlate to the possible knob positions and hence to the states, $\ket{\psi_{j}}{}$, of the object system that are produced. If the die begins in a superposition of $\ket{d_{j}}{}$ states, $\ket{D}{}=\sum_{j=1}^{n}\sqrt{p_{j}}\,\ket{d_{j}}{}$, then the joint state of die and object system following the preparation procedure will be
\begin{equation}
\ket{\Psi}{} = \sum_{j=1}^{n} \sqrt{p_{j}}\,\ket{d_{j}}{}\ket{\psi_{j}}{}. \label{n-c proper}
\end{equation}
This corresponds to the states $\ket{\psi_{j}}{}$ of the object system being mixed with probabilities $p_{j}$; note that the reduced state of the object system indeed takes the form (\ref{proper}).    

At this stage, we may not yet talk of a proper mixture. To do this, further systems need to be introduced. In particular, let us consider adding the environment and an observer.

As we have said, the characteristic feature of a proper mixture is that there is some fact about the state our object system is in, that goes beyond the density operator ascribed to it. In the no-collapse context, such a fact must be understood as relational, that is, as a matter of correlations between the object system, or systems, and states of the environment and observers. Thus we may understand the preparation procedure just outlined as giving rise to a proper mixture if it turns out that following the preparation, the relative state of the object system with respect to the state of some particular observer is one of the states $\ket{\psi_{j}}{}$. For this to happen, the interaction between the systems involved in the preparation procedure and the environment must be such that the observer will become correlated to the die and object system states, or, must be such that an effectively classical record (that is, a record robust against decoherence) of the state of the die and object system is left in the environment. 

Following the preparation of state $\ket{\Psi}{}$ in (\ref{n-c proper}), then, we can imagine two distinct scenarios. In the first, an observer, Alice, indeed becomes correlated to the states produced in the preparation procedure; with respect to her, the object system is in a proper mixture. In the second scenario, another observer, Bob, remains uncorrelated to the states $\ket{d_{j}}{}\ket{\psi_{j}}{}$; with respect to him, the object system is in an \textit{improper} mixture. It is in this sense that the proper/improper mixture distinction becomes relative to the experimental context in no-collapse quantum mechanics.

The proper/improper distinction will be most interesting when we have an ensemble of systems; that is, when the outlined preparation procedure has been run a very large number of times. The total state of the ensemble of object systems and dice will then be of the form: $\ket{\Omega}{}=\ket{\Psi}{}\ket{\Psi}{}\ldots\ket{\Psi}{}\ldots$. If, when an observer is included, his state factorises from $\ket{\Omega}{}$, then with respect to him, the ensemble of object systems will be in an improper mixture (whose state may be written as $\rho^{1}\otimes\rho^{2}\otimes\ldots\otimes\rho^{m}\otimes\ldots$, where $\rho^{i}$ is the density operator of the $i$th object system and is given by the expression (\ref{proper}) ).

However, the state $\ket{\Omega}{}$ may also be written as a superposition of terms which would correspond to the object systems having been prepared in particular sequences of the states $\ket{\psi_{j}}{}$, that is, as a superposition of terms of the form:
\begin{equation}
\ket{d_{j}}{}\ket{\psi_{j}}{}  \ket{d_{j^{\prime}}}{}\ket{\psi_{j^{\prime}}}{} \ldots \ket{d_{j^{\prime\prime}}}{}\ket{\psi_{j^{\prime\prime}}}{}\ldots \label{sequence}
\end{equation}
If an observer becomes \textit{correlated to} (entangled with) these states of the dice and object systems (if, for example, the setting of the apparatus knob for each run of the procedure has left a record in the environment), then with respect to him or her, the ensemble will be in a proper mixture: each of the object systems will be in some particular state $\ket{\psi_{i}}{}$. Of course, the superposition of terms of the form (\ref{sequence}) will include very many sequences in which the relative frequencies of occurrence of the states $\ket{\psi_{j}}{}$ will not match the probabilities $p_{j}$. But, in the usual way \cite{everett,hartle}, as the number of object systems in the ensemble becomes very large, then the overwhelming weight (given by the standard mod squared measure) will lie with those terms in the superposition for which the relative frequency of the appearance of a state $\ket{\psi_{i}}{}$ is very close to its mixing probability $p_{i}$.  

We should note a final important facet of the proper/improper mixture distinction, whether in the context-relative no-collapse quantum mechanical setting, or otherwise. This is the fact that when an ensemble is described as improperly mixed, this means that it is `structureless', or `homogeneous' in von Neumann's sense \cite{vN} (see also \cite{elbybrownfoster}). That is, if the ensemble is divided up into sub-ensembles by some place selection procedure, then the resulting sub-ensembles will have just the same predicted statistics for measurement outcomes as the original ensemble. By contrast, an ensemble that is a proper mixture is not structureless in this way, as we are supposing that there are further facts about the states of each individual system making up the ensemble (even if these facts happen to be construed relationally). In this case, a place selection procedure does in principle exist, that would allow the ensemble to be separated out into statistically distinct sub-ensembles: all that is required is access to these further facts. In picturesque illustration, we might imagine that all along there had been a technician operating the state preparation device, who secretly took a note of the states of the emitted systems, one by one; and who then reveals his list to us. What will be required in the no-collapse context, of course, is access to the dice systems associated with each run of the preparation procedure. Separation into the required sub-ensembles may then proceed via some suitable unitary dynamics conditioned on the dice states associated with each particular object system\footnote{In the no-collapse case, there is the further complication of distinguishing place selection from separation following a measurement interaction. If one is presented with an improper mixture, one can take it into the form of a proper mixture by performing a measurement interaction; this corresponds to the familiar `effective collapse'. One might then proceed to separate the --- now properly mixed --- ensemble. The process of place selection differs from this two-step process, as it is required that the relative states of the object systems with respect to the observer do not change under the sub-ensemble selection procedure. This will not be the case if one first performs a measurement-type interaction in order to change an improper mixture to a proper one before proceeding to select sub-ensembles.}.        

\section{Entanglement}

We now turn to entanglement. The conundrums that are posed by the existence of entangled states in quantum mechanics were first vividly emphasised by Einstein, Podolsky and Rosen \cite{EPR}\footnote{For a recent discussion of the EPR argument and a discussion of aspects of the relationships between entanglement, relativity and nonlocality, see \cite{erpart1}.}. In recent years, with the advent of quantum information theory, it has come to be recognised that entanglement can function as a communication and computational resource; and our understanding of the phenomenon has, accordingly, considerably increased. Achievements include a range of quantitative measures of entanglement and the recognition of qualitatively distinct categories of entangled states (e.g.`bound' versus `distillable' states \cite{horodeckisPRL:1998}, GHZ versus EPR-type entanglement\cite{LindenetalGHZvsEPR}).

A state is called entangled if it is not separable, that is, for bipartite systems, if it cannot be written in the form:\begin{equation}
\rho^{AB}=\sum_{i}\lambda_{i}\rho^{A}_{i}\otimes\rho^{B}_{i}, \label{separable}
\end{equation} 
where $\lambda_{i} \geq 0$, $\sum_{i}\lambda_{i}=1$ and $A$, $B$ label the two subsystems. States of the form (\ref{separable}) are often also called \textit{classically correlated} \cite{werner}, as the outcomes of measurements on these states can always be modelled by a local hidden variable theory (consider in particular the simplest case in which the hidden variable just specifies the quantum states $\rho^{A}_{i}$ and $\rho^{B}_{i}$). For the purposes of \textit{using} entanglement, whether for computation or communication, the overriding question when presented with a pair of systems is whether their state is or is not separable; whether the state can, or cannot be written in the form (\ref{separable})\footnote{A range of operational criteria exist with which to address this question, necessary and sufficient conditions in the $2\otimes 2$ and $2\otimes 3$ cases, and necessary conditions for separability otherwise \cite{horodeckisPLA:1996,peresseparability}. Note that one might also be interested, if the state turns out to be entangled, in whether this entanglement may be distilled and put to use.}. However, if one is also interested in the conceptual questions that entanglement raises, then a more finely grained approach may be desirable.

The point is this: separability is a property of a density operator, and as we know, a given density operator can arise in a variety of different ways. In particular, it is possible that a density operator which is separable can arise from taking a mixture of \textit{entangled} states (we shall see some examples in Section~{\ref{illustrations}}). Thus it appears that the special property of being entangled can be made to disappear simply by taking a mixture of systems.

It is natural to think of the existence of entangled states, ontologically, as corresponding to a distinct feature of holism in the quantum world; entanglement seems to mark the possession by joint systems of properties that differ profoundly from classical physical properties (for example, one might note that the properties of entangled systems are not reducible to properties of the component systems). But, if the entanglement associated with a group of systems can be made to disappear by so innocuous a process as mixing them together, it would seem that entanglement must be too ephemeral a property to be associated with any signficant ontological distinction.

However, we can avoid this rather surprising conclusion by drawing a distinction in the context of entanglement that is analogous to the distinction between proper and improper mixtures discussed above. Note that the problematic scenario arises when one is considering taking a proper mixture of entangled states in such a way that the density operator describing the resultant ensemble is separable. This means, for example, that one will not be able to observe violation of any Bell inequality for measurements on the ensemble.

Imagine, then, that our $\ket{\psi_{j}}{}$ in (\ref{proper}) form a set of entangled states mixed with probabilities $p_{j}$ such that $\rho$ is separable. As the ensemble is a \textit{proper} mixture, then each of the elements of the ensemble will in fact be, by hypothesis, in an entangled state. Since it is a proper mixture and there is a fact about which state each of the elements of the ensemble possesses, there exists, in principle, a selection procedure which would allow us to separate the ensemble back out into sub-ensembles, each of which will be described by a pure state density operator $\ket{\psi_{j}}{}\bra{}{\psi_{j}}$ and which will display all the properties associated with entanglement. Thus, in this case, although the original ensemble is described by a separable density operator, we can nonetheless make sense of there really being entanglement associated with the systems: it hasn't, after all, mysteriously vanished.

\begin{figure}

\begin{center}
\includegraphics{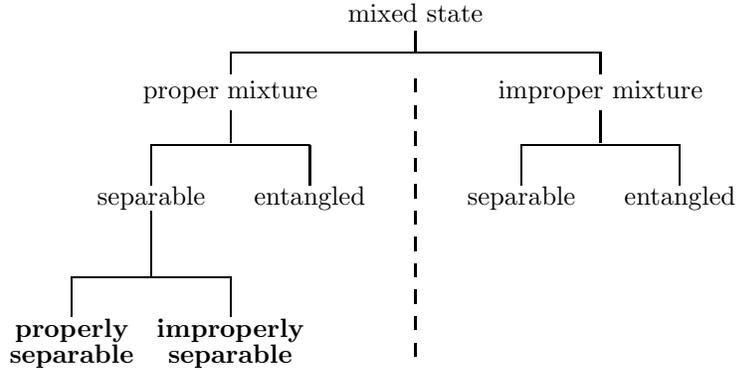}
\end{center}

\caption{\small The place of the proper/improper separability distinction: If one has a \textit{proper} mixture which is separable, it is possible to ask whether the mixture actually contains any entanglement or not. If it does then the mixture is improperly separable, if it does not, the mixture is properly separable. If instead one has an improper mixture, the question of the actual composition of the mixture does not arise. \label{distinction}} 
\end{figure}

We suggest that this sort of scenario, in which a proper mixture containing entangled states gives rise to a separable density operator, be termed an example of \textit{improper separability}. By contrast, we suggest that if a separable density operator results from a proper mixture of \textit{separable} states, then we have a case of \textit{proper separability} (Figure~\ref{distinction}). We see that although entanglement can be made experimentally inaccessible by mixing, this only results in improper, rather than proper separability and there need be no mystery at the conceptual level over the disappearance.

In the no-collapse setting, entanglement is generic: sub-systems into which the world is divided will typically be entangled with one another. Furthermore, we have noted that whether the ensemble resulting from a given preparation procedure may be said to be properly or improperly mixed will depend on the experimental setting. But, given a context in which the ensemble resulting from the preparation procedure can indeed be said to be properly mixed, and if it so happens that the density operator describing this ensemble is a separable one, then the distinction between proper and improper separability may still be applied. The question is whether the object systems making up the ensemble are themselves in entangled states or not. If one simply considers the reduced density operators of the individual object systems, tracing out all other degrees of freedom, then the answer will be in the negative. But, if the ensemble is a proper mixture, we have another way of considering the question. We may ask instead whether the \textit{relative} states of the object systems with respect to the experimentalist are entangled or not. If they are, then the ensemble exhibits improper separability; if not, the ensemble exhibits proper separability. By contrast, if we focus on a context with respect to which our ensemble is in an improper mixture, then this question about the relative states of the systems making up the ensemble doesn't arise. One can then only enquire about the reduced states of each individual system, which in this case will be separable.

\section{Illustrations}\label{illustrations}

The fact that mixtures of entangled states can give rise to separable states is a commonplace in entanglement theory, but that it may be of conceptual importance has so far been little noted, with the exception of some remarks by Popescu and Collins \cite{popescuPRL:1994,collins:popescu} and Seevinck and Uffink \cite{seevinckuffink}. We should perhaps not be too surprised that this phenomenon may occur. Consider the four Bell-states,
\[\begin{array}{c}
\ket{\phi^{+}}{}=1/\sqrt{2}(\ket{\up}{}\ket{\up}{}+\ket{\down}{}\ket{\down}{}),\\
\ket{\phi^{-}}{}=1/\sqrt{2}(\ket{\up}{}\ket{\up}{}-\ket{\down}{}\ket{\down}{}),\\
\ket{\psi^{+}}{}=1/\sqrt{2}(\ket{\up}{}\ket{\down}{}+\ket{\down}{}\ket{\up}{}),\\
\ket{\psi^{-}}{}=1/\sqrt{2}(\ket{\up}{}\ket{\down}{}-\ket{\down}{}\ket{\up}{}).
\end{array} \]
These maximally entangled states form an orthonormal basis for $2\otimes2$ dimensional quantum systems, hence projectors onto these states satisfy the completeness relation, summing to the identity operator $\mathbf{1}\otimes\mathbf{1}$; thus an equally weighted mixture of these states will correspond to the maximally mixed state $1/4(\mathbf{1}\otimes\mathbf{1})$, which, of course, is separable. Less trivial examples exist. We shall consider, in illustration, what happens when one takes general convex combinations of projectors onto the Bell states, giving rise to the class of so-called \textit{Bell-diagonal} states\footnote{The Bell-diagonal states and their $U_{1}\otimes U_{2}$ equivalents are a proper subset of the set of $2\otimes 2$ density operators, as mixtures of maximally entangled states always have maximally mixed reduced states for subsystems, which will not be the case for general $2\otimes 2$ density operators, whether entangled or separable.}.

\begin{figure}
\begin{center}\scalebox{0.8}{\includegraphics{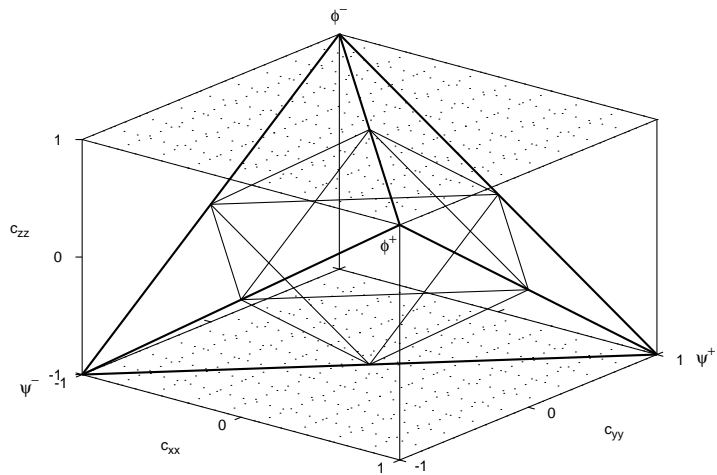}}\end{center}
\caption{\small Bell-diagonal states may be represented by the diagonal components of the correlation matrix $c_{ij}$. The vertices of the tetrahedron $\cal{T}$ correspond to the four Bell states $\ket{\phi^{+}}{},\ket{\phi^{-}}{},\ket{\psi^{+}}{},\ket{\psi^{-}}{}$. A Bell-diagonal state is separable iff it corresponds to a point belonging to the central octohedron $\cal{T}\cap-\cal{T}$.\label{tetrahedron}}
\end{figure}

It is often useful to consider the set of Hermitian operators on an $n$-dimensional complex Hilbert space as themselves forming a real Hilbert space of $n^{2}$ dimensions, on which we have defined a scalar product $(A,B)=\mathrm{Tr}(AB)$ \cite{Fano}. For $n=2^{m}$, $m$-fold tensor products of the Pauli operators and the identity constitute a convenient basis set for this space. The density operator for an $n$-dimensional quantum system can then be written as a vector whose (real) components are simply the expectation values of these basis operators. In particular, the density operator for a $2\otimes2$ dimensional system may be written in the general form:
\begin{equation}
\rho^{AB}=\frac{1}{4}\Bigl( \mathbf{1}\otimes\mathbf{1} + \mathbf{a}.\boldsymbol{\sigma}\otimes\mathbf{1} + \mathbf{1}\otimes\mathbf{b}.\boldsymbol{\sigma} +\sum_{ij}c_{ij}\sigma_{i}\otimes\sigma_{j}\Bigr),
\end{equation}      
where $a_{i}$, $b_{i}$ and $c_{ij}$ are the expectation values of the operators $\sigma_{i}\otimes\mathbf{1}$, $\mathbf{1}\otimes\sigma_{i}$ and $\sigma_{i}\otimes\sigma_{j}$, respectively. For the projectors onto the four Bell states, $\mathbf{a}=\mathbf{b}=0$, and $c_{ij}$ is diagonal. So in this case, we need only consider the possible values of the diagonal components $c_{ii}$, which will form a vector in a 3-dimensional real space, allowing one to represent the states in an easily visualisable manner. As is well known, the vectors corresponding to the Bell state projectors are:
\begin{equation}
\mathbf{c}_{\phi +}= \begin{pmatrix} 1 \\ -1 \\ 1\end{pmatrix}, \mathbf{c}_{\phi -}= \begin{pmatrix} -1 \\ 1 \\ 1\end{pmatrix}, \mathbf{c}_{\psi +}= \begin{pmatrix} 1 \\ 1 \\ -1\end{pmatrix}, \mathbf{c}_{\psi -}= \begin{pmatrix} -1 \\ -1 \\ -1\end{pmatrix}.  
\end{equation}  
These four vectors correspond to the vertices of a tetrahedron $\cal{T}$ centred on the origin (see Figure~\ref{tetrahedron}); the Bell-diagonal states, given by convex combinations of the Bell-state projectors, will then correspond to vectors lying on or within the surfaces of $\cal{T}$. It has been shown \cite{horodeckisPRA:1996} that a Bell-diagonal state is separable if and only if the end point of its corresponding vector $\mathbf{c}$ lies on or within the octohedron given by the intersection of $\cal{T}$ with its reflection through the origin, $-\cal{T}$ (Figure~\ref{tetrahedron}).

We now have a very clear picture of when mixing maximally entangled states of a $2\otimes2$ system can give rise to (improperly) separable states (note that degree of entanglement is invariant under $U_{1}\otimes U_{2}$ rotation, hence an identical picture may be drawn for any maximally entangled basis set): it will happen whenever the mixing probabilities take us from the vertices of the tetrahedron into the central separable octohedron. In particular, let us consider the example of mixing $\ket{\phi^{+}}{}$ and $\ket{\phi^{-}}{}$. In this case, we will be constrained to the plane $c_{zz}=1$ and mixing the two states will move us along the line from $(c_{xx},c_{yy})=(-1,1)$ to $(c_{xx},c_{yy})=(1,-1)$. The octohedron of separable states only intersects the plane $c_{zz}=1$ at the single point above the origin, $(c_{xx},c_{yy})=(0,0)$, i.e. when we have an exact 50/50 mixture of the two entangled states. The resultant state is the (improperly) separable state with maximal classical correlations: $\rho =1/2(\ket{\up\up}{}\bra{}{\,\up\up\!}+ \ket{\down\down}{}\bra{}{\,\down\down\!})$.\footnote{A related example has also been given by Seevinck and Uffink \cite{seevinckuffink} in the context of distinguishing genuine $N$-party entanglement from $M<N$-party entanglement. They note that an $N$-party state can be $M$-party entangled even though there is no $M$-party subsystem whose reduced state is $M$-party entangled. The example they give is precisely an example of improper separability, in which the 2-party entanglement of the pair of Bell states $\ket{\psi^{+}}{}$ and $\ket{\psi^{-}}{}$ becomes hidden on mixing.}

As another familiar example, consider the generalized Werner states for $2\otimes2$ systems. These are mixtures of the singlet state $\ket{\psi^{-}}{}$ and spherically symmetric noise, and take the form\[
\rho(\lambda)= \lambda \ket{\psi^{-}}{}\bra{}{\psi^{-}} + (1-\lambda)\frac{1}{4}\mathbf{1}\otimes\mathbf{1} = \frac{1}{4}\bigr(\mathbf{1}\otimes\mathbf{1} - \lambda(\sigma_{x}\otimes\sigma_{x}+ \sigma_{y}\otimes\sigma_{y} + \sigma_{z}\otimes\sigma_{z})\bigl).\]   
These states lie on the line joining the origin to the vertex representing the singlet state.
The class is of particular interest as for certain ranges of the value of $\lambda$, the state will be entangled, but will not violate any Bell inequality, a possibility first noted by Werner \cite{werner}. The mixture will move from being separable to entangled when it crosses the surface of the octohedron, a distance of $1/\sqrt{3}$ along the line from the origin, corresponding to $\lambda=1/3$. (It follows from a result of the Horodeckis giving necessary and sufficient conditions for Bell inequality violation in the case of two qubits \cite{horodeckisPLA:1995} that the mixture will not violate a Bell inequality until the distance is greater than $\sqrt{3}/\sqrt{2}$ from the origin, $\lambda >1/\sqrt{2}$.) Mixing the singlet state with a fraction $>2/3$ of the maximally mixed state (which, as we have said, may itself be an equal mixture of the four Bell states) will thus result, again, in an improperly separable state. 

Reflecting on the case of Werner states raises the following possibility. One could imagine taking a proper mixture of states which are themselves entangled but do not violate any Bell inequality (hence are necessarily mixed \cite{popescurohrlich}, but perhaps improperly), in such a way that the resulting density operator is separable. This would be another example of improper separability. However, in this case, although the possibility exists in principle of selecting out sub-ensembles corresponding to each of the states making up the mixture, these sub-ensembles will not make their entanglement manifest by violating a Bell inequality. Popescu has shown \cite{popescu:hidden} that for dimensions greater than five, Werner states can be made to display `hidden' non-locality, i.e., violate a Bell inequality after a sequence of measurements, but this particular technique will not be applicable with dimensions $2\otimes2$. It would thus seem that when improper separability results from mixing entangled states that do not themselves violate any (at least single measurement) Bell inequality, although it remains true that the ensemble can be said genuinely to contain some entanglement at the ontological level, it is certainly lying very low.

\section{Discussion}
We began by noting that the venerable distinction between proper and improper mixtures may be sustained in the context of no-collapse versions of quantum mechanics, if it is recognised that the distinction becomes relative to the experimental context.
We then remarked on the fact that the process of mixing entangled states, even \textit{maximally} entangled states, can, perhaps somewhat surprisingly, result in separable states. The examples given by the Bell diagonal states illustrate how this can happen and that the phenomenon is widespread.  This would appear to make entanglement a disturbingly ephemeral property at the ontological level. However, by introducing a distinction analogous to that between proper and improper mixtures, a distinction between proper and improper separability, we have seen that it remains possible to retain an ontologically robust notion of entanglement.    

\section*{Acknowledgements}

We would like to thank Jon Barrett and Michiel Seevinck for discussion and also the anonymous referee for 
comments which resulted in clarification of some key arguments in the paper.


{\small     
\begin{thebibliography}{10}

\bibitem{d'Espagnat}
B~d'Espagnat.
\newblock {\em Conceptual Foundations of Quantum Mechanics}.
\newblock Addison-Wesley, second edition, 1976.

\bibitem{schrodinger2}
Erwin Schr{\"o}dinger.
\newblock Probability relations between separated systems.
\newblock {\em Proc. Camb. Phil. Soc.}, 32:446--452, 1936.

\bibitem{jaynes:1957ii}
Edwin~T Jaynes.
\newblock Information theory and statistical mechanics {II}.
\newblock {\em Phys. Rev.}, 108:171, 1957.

\bibitem{hjw}
L~P Hughston R Jozsa W~K Wootters.
\newblock A complete classification of quantum ensembles having a given density
  matrix.
\newblock {\em Phys. Lett. A}, 183:14--18, 1993.

\bibitem{decoherence}
Wojciech~Hubert Zurek.
\newblock Decoherence, einselection, and the quantum origins of the classical.
\newblock {\em Rev. Mod. Phys.}, 75(3):715--775, 2003.

\bibitem{everett}
Hugh Everett, III.
\newblock ``{R}elative state" formulation of quantum mechanics.
\newblock {\em Rev. Mod. Phys.}, 29:454--62, 1957.

\bibitem{simonrelativism}
Simon Saunders.
\newblock Relativism.
\newblock In R.~Clifton, editor, {\em Perspectives on Quantum Reality}, pages
  125--142. Kluwer Academic Publishers, Dordrecht, 1996.

\bibitem{Kirkpatrick}
K~A Kirkpatrick.
\newblock Indistinguishability and improper mixtures, 2001.
\newblock ar{X}iv:quant-ph/0109146.

\bibitem{d'Espagnatreply}
B~d'Espagnat.
\newblock Reply to {K} {A} {K}irkpatrick, 2001.
\newblock ar{X}iv:quant-ph/0111081.

\bibitem{hartle}
J~B Hartle.
\newblock Quantum mechanics of individual systems.
\newblock {\em Am. J. Phys.}, 36:704--712, 1968.

\bibitem{vN}
J~von Neumann.
\newblock {\em Mathematical Foundations of Quantum Mechanics}, chapter~4.
\newblock Princeton University Press, 1955.
\newblock English Translation.

\bibitem{elbybrownfoster}
Andrew Elby, Harvey~R Brown, and Sara Foster.
\newblock What makes a physical theory ``complete"?
\newblock {\em Foundations of Physics}, 23(7), 1993.

\bibitem{EPR}
A.~Einstein, B.~Podolsky, and N.~Rosen.
\newblock Can quantum mechanical description of physical reality be considered
  complete?
\newblock {\em Phys. Rev.}, 47:777, 1935.

\bibitem{erpart1}
Christopher~G. Timpson and Harvey~R. Brown.
\newblock Entanglement and relativity.
\newblock In Rosella Lupacchini and Vincenzo Fano, editors, {\em Understanding
  Physical Knowledge}. University of Bologna, CLUEB, Bologna, 2002.
\newblock ar{X}iv:quant-ph/0212140.

\bibitem{horodeckisPRL:1998}
Michal Horodecki, Pawel Horodecki, and Ryszard Horodecki.
\newblock Mixed-state entanglement and distillation: Is there a ``bound"
  entanglement in nature?
\newblock {\em Phys. Rev. Lett.}, 80:5239, 1998.

\bibitem{LindenetalGHZvsEPR}
N~Linden, S~Popescu, B~Schumacher, and M~Westmoreland.
\newblock Reversibility of local transformations of multiparticle entanglement,
  1999.
\newblock ar{X}iv:quant-ph/9912039.

\bibitem{werner}
Reinhard~F. Werner.
\newblock Quantum states with {E}instein-{P}odolsky-{R}osen correlations
  admitting a hidden-variable model.
\newblock {\em Phys. Rev. A}, 40(8):4277--4281, 1989.

\bibitem{horodeckisPLA:1996}
Michal Horodecki, Pawel Horodecki, and Ryszard Horodecki.
\newblock Separability of mixed states: Necessary and sufficient conditions.
\newblock {\em Phys. Lett. A}, 223, 1996.

\bibitem{peresseparability}
Asher Peres.
\newblock Separability criterion for density matrices.
\newblock {\em Phys. Rev. Lett.}, 77(8):1413--1415, 1996.

\bibitem{popescuPRL:1994}
Sandu Popescu.
\newblock Bell's inequalities versus teleportation: What is nonlocality?
\newblock {\em Phys. Rev. Lett.}, 72(6):797--799, 1994.

\bibitem{collins:popescu}
Daniel Collins and Sandu Popescu.
\newblock Classical analog of entanglement.
\newblock {\em Phys Rev A}, 65(3):032321, 2002.
\newblock ar{X}iv:quant-ph/0107082.

\bibitem{seevinckuffink}
Michael Seevinck and Jos Uffink.
\newblock Sufficient conditions for three-particle entanglement and their tests
  in recent experiments.
\newblock {\em Phys. Rev. A}, 65:012107, 2001.

\bibitem{Fano}
U.~Fano.
\newblock Description of states in quantum mechanics by density operator
  techniques.
\newblock {\em Rev. Mod. Phys.}, 29(1):74--93, 1957.

\bibitem{horodeckisPRA:1996}
Ryszard Horodecki and Michal Horodecki.
\newblock Information theoretic aspects of inseparability of mixed states.
\newblock {\em Phys. Rev. A}, 54(3):1838--1843, 1996.

\bibitem{horodeckisPLA:1995}
R~Horodecki, P~Horodecki, and M~Horodecki.
\newblock Violating {B}ell inequality by mixed spin-1/2 states: Necessary and
  sufficient conditions.
\newblock {\em Phys. Lett. A}, 200:340--344, 1995.

\bibitem{popescurohrlich}
S.~Popescu and D.~Rohrlich.
\newblock Generic quantum nonlocality.
\newblock {\em Phys. Lett. A}, 166:293--297, 1992.

\bibitem{popescu:hidden}
S.~Popescu.
\newblock Bell's inequalities and density matrices. {R}evealing `hidden'
  non-locality.
\newblock {\em Phys. Rev. Lett.}, 74:2619--2622, 1995.
\newblock {a}r{X}iv:quant-ph/9502005.

\end{thebibliography}
}
\end{document}